\documentclass[apj,numberedappendix]{emulateapj}

\newcommand{\beq}{\begin{equation}}
\newcommand{\eeq}{\end{equation}}
\newcommand{\beqa}{\begin{eqnarray}}
\newcommand{\eeqa}{\end{eqnarray}}

\newcommand{\mpc}{\ensuremath{{\rm\,Mpc}}}

\newcommand{\hmpc}{\ensuremath{h^{-1}{\rm\,Mpc}}}
\newcommand{\ihmpc}{\ensuremath{h{\rm\,Mpc}^{-1}}}

\newcommand{\hgpcC}{\ensuremath{h^{-3}{\rm\,Gpc^3}}}

\newcommand{\hmsun}{\ensuremath{h^{-1}\ {\rm M_\odot}}}

\newcommand{\Mpc}{\mpc}

\begin{document}

\title{Improving Cosmological Distance Measurements by \\
Reconstruction of the Baryon Acoustic Peak}
\author{Daniel J.\ Eisenstein\altaffilmark{1},
Hee-Jong Seo\altaffilmark{1}, Edwin Sirko\altaffilmark{2}, 
and David N.\ Spergel\altaffilmark{2}}  

\altaffiltext{1}{Steward Observatory, University of Arizona,
                933 N. Cherry Ave., Tucson, AZ 85121}
\altaffiltext{2}{Dept.\ of Astrophysical Sciences, Peyton Hall,
	Princeton University, Princeton, NJ 08544-1001}

\submitted{Submitted to the {\it Astrophysical Journal} April 17, 2006}

\begin{abstract}
The baryon acoustic oscillations are a promising route to the 
precision measure of the cosmological distance scale and hence
the measurement of the time evolution of dark energy.
We show that the non-linear degradation of the acoustic signature
in the correlations of low-redshift galaxies is a correctable process.  
By suitable reconstruction of the linear density field, one can
sharpen the acoustic peak in the correlation function or,
equivalently, restore the higher harmonics of the oscillations 
in the power spectrum.  With this, one can achieve better measurements
of the acoustic scale for a given survey volume. 
Reconstruction is particularly effective at low redshift, where
the non-linearities are worse but where the dark energy density
is highest.  At $z=0.3$, we find that one can 
reduce the sample variance error bar on the acoustic scale 
by at least a factor of 2 and in principle by nearly a factor of 4.
We discuss the significant implications our results have for the 
design of galaxy surveys aimed at measuring the distance scale 
through the acoustic peak.
\end{abstract}

\keywords{large-scale structure of the universe
  ---
  distance scale
  ---
  cosmological parameters
  ---
  cosmic microwave background
}
\maketitle

\section{Introduction}

The late-time acceleration of the expansion rate of the Universe
\citep{Rie98,Per99,Rie04}
argues for a remarkable change to our understanding of the forces
of nature.  Choosing betwen the exotic explanations to this surprising
phenomenon may be possible through precision measure
of the expansion rate of the Universe and of the growth of 
cosmological structure over time.  Sound waves propagating in
the first 400,000 years after the Big Bang produce a characteristic
length scale in the anisotropies of the microwave background and
in the clustering of galaxies \citep{Pee70,Sun70,Bon87,Hol89,Hu96,Eis98}.  
CMB experiments have clearly detected these fluctuations in both
the temperature and polarization power spectrum \citep[for a summary of 
recent results, see][]{Spe06}.
With large galaxy surveys, we
can detect this acoustic signature as a peak in the correlation
function at $\sim 150 \Mpc$ or as a harmonic sequence of oscillations
in the power spectrum \citep{Col05,Eis05}.  
This length scale can be measured as a
characteristic angle on the sky, yielding the angular diameter
distance as a function of redshift, and as a characteristic
difference in redshift of galaxy pairs along the line of sight,
yielding the Hubble parameter $H(z)$ \citep{Eis03,Bla03,Hu03,Seo03}.  
Because of the very large
scale, the acoustic signature remains in the linear regime even
today and is therefore a highly robust method for measuring the 
cosmological distance scale \citep{Mei99,Seo05,Eis06}.

However, as the universe evolves, the acoustic signature in the
correlation function is broadened by non-linear effects.  Non-linear
gravitational structure formation and redshift distortions move galaxies
away from their original locations, blurring out the peak at 150 Mpc
separation, or, equivalently, erasing the higher harmonics in the power
spectrum that represent smaller scales \citep{Eis06}.  Even though the
baryon bump remains \emph{measurable} in the correlation function, these
nonlinear effects reduce the \emph{precision} of the distance scale that
can be measured from a given volume of space, by roughly a factor of
three at the present day.  This loss is significant because survey volume
is precious: larger surveys are more expensive and the amount of volume
that we {\it can} survey in the low-redshift Universe is limited.

We argue in this paper that this loss of precision is avoidable.
The blurring of the acoustic peak is largely due to bulk flows
and super-cluster formation, effects that are generated by gravitational
forces on large scales.  The same map of galaxies intended
to measure the acoustic scale is an accurate map of the large-scale gravitational
source terms.  One can essentially run the gravitational flow backwards
to restore the acoustic peak nearly to its linear regime shape.

\section{Non-linearities}

\begin{figure}[t]
\plotone{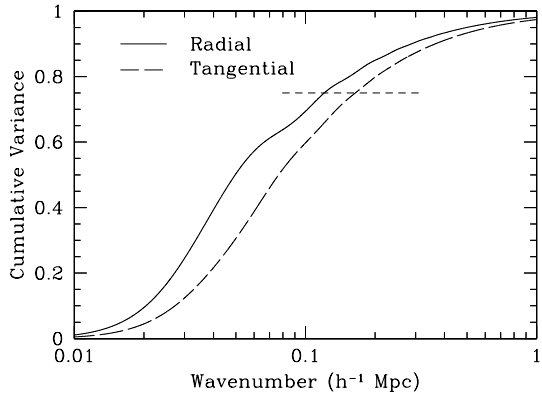}
\caption{\label{fig:zel}%
The cumulative variance in the differential motion of pairs 
initially separated by 150 Mpc as a function of cutoff wavenumber,
normalized to the total variance.
These curves are calculated in the Zel'dovich approximation according
to the formulae in \citet{Eis06}.  The displacement along the separation
vector is shown as the solid line; the displacement along a single 
direction perpendicular to the separation vector is shown as the dashed
line.  Most of the integral is supported by wavenumbers between 
0.02 and 0.2$\ihmpc$.  The horizontal dashed line is drawn at 75\%,
where half of the rms displacement has been fixed.
}
\end{figure}

Numerical simulations show that the advancing scale of non-linear
gravitational collapse erases the higher harmonics of the acoustic
oscillations \citep{Mei99,Seo05,Spr05,Whi05}.  
\citet{Eis05,Eis06} note that the harmonic sequence in the power
spectrum correspond to a single peak in the correlation function
and that the damping envelope corresponds to the broadening of this
peak.  In the configuration-space view, the source of the late-time 
broadening is clear: matter is being moved by roughly 10 Mpc from its initial
position.  \citet{Eis06} build a model for the non-linearity in terms
of the differential motion of pairs initially separated by 150 Mpc.
The final large-scale correlation function is simply the convolution of
the linear correlation function with the distribution of differential
motions.

One can ask what scales are responsible for the differential motion.
Using the Zel'dovich approximation \citep{zeldovich_1970}, the second moment
of the distribution of the differential motion can be written as an integral over the 
power spectrum \citep{Eis06}.  The cumulant of the integrand is plotted 
in Figure \ref{fig:zel}.  Here we see that most of the motion is
generated at $k\approx0.1\ihmpc$.  Wavenumbers smaller than $0.02\ihmpc$
contribute little, because these perturbations affect both points 
equally.  Wavenumbers larger than $0.2\ihmpc$ contribute less because
the CDM power spectrum is fairly red on these scales.  

The large-scale velocity field is responsible for most of the nonlinear effects
that appear to ``erase" the high order acoustic peaks in the power spectrum.
 The motions are dominantly due to bulk flows
and the formation of superclusters.  Small-scale formation of halos
is subdominant; even big halos only pull material from an average
of $5(M/10^{14}\hmsun)^{1/3} \hmpc$ away, about half the bulk flow
motions.

\section{Reconstruction}

Our major point is that because the scales of interest are large,
the motions of galaxies can be well modeled by perturbation theory and in principle
can be measured and removed.  In the simplest terms, the bulk flows
are generated by exactly the density perturbations that are being
surveyed to measure the acoustic oscillations.  The connection 
between the density and velocity field on these large scales is
nearly that of linear theory, so one can predict the velocity
field and undo the motion of the galaxies.

In more detail, reconstruction of the velocity field or the
linear density field is a subject with considerable history.
\citet{peebles_1989,peebles_1990} pioneered the subject by reconstructing
the trajectories of Local Group galaxies using the principle of
least action, with the goal of constraining the local value of $H_0$.
The simple ``Gaussianization'' method of \citet{weinberg_1992} approached
the problem from a cosmological perspective.  \citet{nusser_dekel_1992}
and \citet{gramann_1993} reconstructed primordial density fields with
a technique based on the success of the Zeldovich approximation in
the quasilinear regime, similar to our technique described below.
\citet{narayanan_croft_1999} developed these techniques further (and
offered a comparison of several different methods). 
\citet{monaco_efstathiou_1999} investigated a self-consistent, but more complicated, iterative scheme in which the initial densities source the Lagrangian map, rather than the final densities.  Pursuit of this reconstruction method further may warrant using higher order relationships between peculiar velocities and densities \citep[e.g.,][]{chodorowski_etal_1998}.  The techniques of
\citet{croft_gaztanaga_1997} and \citet{brenier_etal_2003} are based on
the principle of least action again, with the approximation that particles
move on straight trajectories.  As far as we know no one has attempted
a full cosmologically relevant proper least action integration backwards
in time, as the computational resources would be prohibitively expensive.
Reversing the sign of gravity and running an evolved simulation backwards
would not work, at least in the realistic case in which there are errors
or unconstrained dimensions in the galaxies' phase space, because the
decaying modes would overpower the growing modes as one went back in time.

Restoring in full the acoustic signature at $k<0.2\ihmpc$ is 
an undemanding application of these reconstruction techniques.
To demonstrate that reconstruction can help, we here show a 
simple method.  We take the $z=0.3$ outputs from the $512\ \hmpc$ box N-body simulations 
from \citet{Seo05} ($N = 256^3$ particles), compute the density field, Fourier transform
and filter with a Gaussian of $10\hmpc$ or $20\hmpc$ width.  From this, we predict
the linear theory motion using $\nabla\cdot\vec{q} = -\delta$,
where $\vec q$ is the Lagrangian displacement field and $\delta$ is the 
fractional overdensity.  We then move the particles by $-\vec q$.
We do the same for a reference grid of smoothly distributed particles.
In redshift space, we account for the linear redshift distortions \citep{Kai87} by
boosting the displacements along the line of sight
by an additional factor of $1+f$, where $f=d(\ln D)/d(\ln a)$, $D$ is the 
linear growth function, and $a$ is the scale factor of the Universe,
A new density field is defined by the difference of the density field of 
the real particles and that of the reference particles.  
Note that, in contrast to the work of \citet{sirko_etal_2006},
the point is not to move all of the particles back to
their initial location, but
rather to move the measured densities back to their initial location.  

\begin{figure}[t]
\plotone{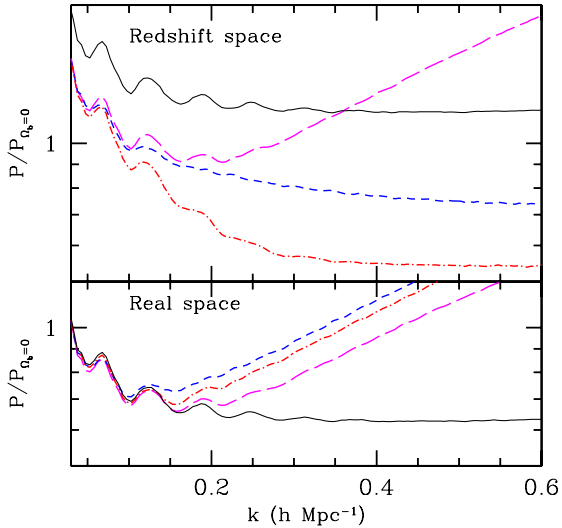}
\caption{\label{fig:recon}%
The matter power spectrum after reconstruction by the linear-theory
density-velocity relation, with the density field Gaussian filtered.
The bottom panel shows the real-space power spectrum; 
the top panel shows the spherically averaged redshift-space power spectrum.
The black solid line shows the input power spectrum
at $z=49$; it has been displaced in the top panel for clarity.  
The blue short-dashed line shows the matter power spectrum at $z=0.3$;
one can see that acoustic peaks have been lost.  
In the bottom panel, 
the red dot-dashed line and magenta long-dashed line
show the effects of reconstruction for $20\hmpc$ and $10\hmpc$
Gaussian filtering, respectively.  
In the top panel, both lines show $10\hmpc$ filtering; the 
red dot-dashed line is without finger of God compression, while
the magenta long-dashed line includes compression.
The increase of power at large wavenumbers is essentially irrelevant to 
the quality of the acoustic signature; one would in practice marginalize
over these broadband changes.
}
\end{figure}

\begin{figure}[t]
\plotone{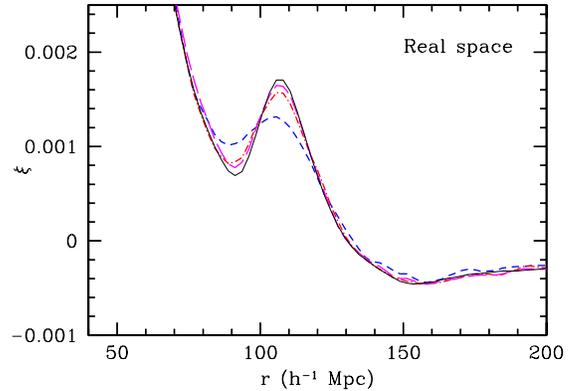}
\caption{\label{fig:xireal}%
The real-space matter correlation function after reconstruction by the linear-theory
density-velocity relation, with the density field Gaussian filtered.
The black solid line shows the correlation function at $z=49$.  
The blue 
short-dashed line shows it at $z=0.3$; 
the acoustic peak has been smeared out.
The red dot-dashed and magenta long-dashed 
lines show the effects of reconstruction for $20\hmpc$ and $10\hmpc$
Gaussian filtering, respectively.  Even this very simple reconstruction
recovers nearly all of the linear acoustic peak.
}
\end{figure}

\begin{figure}[t]
\plotone{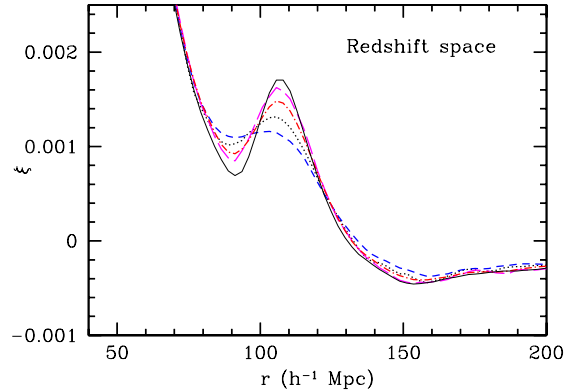}
\caption{\label{fig:xired}%
The redshift-space matter correlation function after reconstruction by the linear-theory
density-velocity relation, with the density field Gaussian filtered.
The black solid line shows the correlation function at $z=49$.  
The blue short-dashed line shows the redshift-space correlation function at $z=0.3$; 
the acoustic peak has been smeared out.
The black dotted line shows the real-space correlation function for comparison.
The red dot-dashed line line shows the effects of reconstruction for a $10\hmpc$
Gaussian filtering; the magenta long-dashed line is the result when one compresses
the fingers of God prior to the reconstruction.
These reconstructions significantly improve the acoustic peak.
}
\end{figure}

The power
spectrum of this field is shown in Figure \ref{fig:recon}.
This Figure is the average of power spectra from 30 simulations at
$z=0.3$.  One can see the degradation of the higher harmonics
at $z=0.3$ in the uncorrected density field, compared to the initial spectrum at $z=49$, and see that the reconstruction has partially restored them.
Figures \ref{fig:xireal} and \ref{fig:xired} show the correlation
functions in real and redshift space.  The acoustic peak is nearly
fully restored in real space.  In redshift space, the peak is considerably
improved but not fully fixed.  

When one predicts the large-scale displacement field,
one is also predicting the large-scale velocity field and hence
the correction for large-scale peculiar velocities in redshift space.
This is important because \citet{Eis06} find that redshift space
distortions degrade the radial measurement of the acoustic peak.
On large scales, the real-space displacements of particles are in
the same direction as their peculiar velocity distortion, so the
degradation of the acoustic peak is worse in redshift space than 
in real space.  Reconstruction can fix this.  

However, there are also redshift-space distortions from small-scale
peculiar velocities, i.e., fingers of God.  Clusters of
galaxies appear as long cigars along the line of sight in redshift space.
For the purposes of determining bulk flows, one should simply compress
these fingers of God back to some approximation of their real-space
location.  Without this step, the fingers of God get stretched out 
further by the reconstruction, degrading the acoustic signature.
To show that finger-of-God compression can help the reconstruction,
we have identified clusters in redshift space with an anisotropic
friends-of-friends algorithm and moved all cluster particles to the
center of mass of the cluster.  The correlation function that results
from running our reconstruction on the compressed density field is
shown in Figure \ref{fig:xired}.  One sees a modest but useful improvement.

The simple reconstruction described above has not fully restored
the linear acoustic scale, particularly when beginning from redshift
space.  We expect that more sophisticated reconstruction methods will produce
further improvements.  The small end of the range of the scales of
interest are in the quasi-linear regime, and our assumption of linear
theory for both the continuity equation and the redshift distortions
is only a first approximation here.  

We use a Fisher matrix calculation to estimate how much the reconstruction
has improved the recovery of the acoustic scale.  Our calculation is
based on the methods in \citet{Seo03} but with the derivatives multiplied
by a Gaussian filtering that is tuned to match the pair-wise Lagrangian
displacement \citep{Eis06} and that visually reproduces the smearing of
the acoustic peak.  We focus here on the spherically-averaged acoustic
scale; we will present anisotropic results (i.e., separate estimates for
the angular diameter distance and the Hubble parameter) in a future paper.
We find that at $z=0.3$, in the absence of reconstruction, a survey
of $1\hgpcC$ should produce a distance measurement of 1.4\% from real
space and 1.9\% from redshift space.  The real-space value is in good
agreement from the jack-knife estimates from the simulated real-space
power spectra in \citep{Seo05}.  The simple reconstruction presented in this
section improves these measurements to 0.75\% and 0.95\% in real and
redshift space, respectively.  This is a factor of two improvement.
We find that if the reconstruction were perfect, one could achieve 
0.5\% distances (although shot noise from reasonable galaxy samples 
would degrade this to 0.55--0.60\%).  Hence, the raw precision at $z=0.3$
is roughly a factor of 3.5 worse than the true cosmic variance, and 
the simplest reconstruction returns the first factor of two.
At higher redshifts, the cosmic variance per unit volume won't change,
but the data sets with or without reconstruction will be closer to that
ideal.

\section{Discussion}

We have demonstrated that density-field reconstruction can restore the
linear-regime contrast of the baryon acoustic signature.  Physically,
the galaxy map locates the superclusters and voids and permits one to
predict the large-scale flows resulting from these objects.  By moving
objects back to their initial location, one can correct the effect
that these flows have on the characteristic separation of galaxies
produced by the acoustic waves in the early universe.  By restoring the
linear-regime clustering, one can improve the precision of the measurement
of the acoustic scale from galaxy redshift surveys and hence the available
constraints on the cosmic distance scale and dark energy.

The opportunity for density reconstruction alters the optimization
of surveys for baryon acoustic oscillations.  Low-redshift surveys
were thought to suffer in performance per unit volume; with 
reconstruction, this is not the case.  Given that dark energy
is more important (and hence easier to measure) at lower redshift,
this will tend to push the optimal redshift range lower.  On the
other hand, lower redshift surveys were generally designed at lower
target densities, because the suppression of the higher harmonics
meant that it was not necessary to measure $k\approx0.2\ihmpc$
well.  Now the opportunity exists to get acoustic oscillation
information from the higher harmonics, so that one might want 
to reduce shot noise and measure $k\approx0.2\ihmpc$ well.
In practice, with multiobject spectrographs, the number density
at the highest redshifts will drop due to the flux limits imposed
by a given exposure time, but at lower redshifts one can use the
extra galaxies available to that flux limit, if one has sufficient
number of fibers or slits.

Do the redshift surveys aimed at acoustic oscillations have enough
information to do density-field reconstruction?  Acoustic oscillation
surveys are designed to balance shot noise and sample variance at the
wavenumbers where the acoustic oscillations are found, namely $0.1-0.2\ihmpc$.
If a survey is measuring at $k\lesssim0.2\ihmpc$ with shot noise 
below sample variance, then it is producing a reasonable fidelity
map at exactly the wavenumbers required to do the reconstruction,
as shown in Figure \ref{fig:zel}.
Turning this around, if the map is insufficient for reconstruction,
it will also be too noisy to measure the power spectrum well
(i.e., approaching the sample variance limit) at the higher harmonics
that one was hoping to improve.  
Indeed, within the approximation of the linear-theory density-displacement 
relation, the desired weighting of each Fourier mode to be used in
reconstruction is $nP/(1+nP)$, where $n$ is the comoving number density
and $P$ is the power spectrum at that wavenumber.  This is the 
same familiar factor that determines the measurement of modes in the
power spectrum \citep{Fel94,Teg97}.
Hence, surveys designed to measure
the acoustic oscillations with moderate signal-to-noise ratio maps
will be well suited to reconstruction.  


One must remember that the reconstruction need not be perfect.  
The propagation of sound waves at the epoch of recombination 
has mild dispersion that gives the linear regime acoustic peak a
width of about $30\mpc$ FWHM or $8\hmpc$ rms width, comparable
to the pairwise displacments of galaxies even at low redshifts \citep{Eis06}. Once we reduce the errors below this characteristic width, there is little improvement in the acoustic oscillations
measurements.   Thus, reconstruction is  easier at higher
redshifts: the raw displacements are smaller, so one can accomplish
sufficient reconstruction using only larger scales.  Even at 
$z=3$, it is likely that reconstruction will benefit the recovery
of the line-of-sight acoustic peak (and hence $H(z)$), because
\citet{Eis06} shows that the redshift-space displacements are
nearly double the real-space ones at $z\gtrsim1$ and hence are
not quite negligible even at $z=3$.

Reconstruction places a premium on surveys with contiguous area.
Surveys with lots of gaps on 10-100 Mpc scales will not measure the
density field well enough to do these non-linear corrections: the
source of bulk flows could be hidden in the gaps.  Some gaps, e.g., due
to bright stars, are inevitable and will degrade the reconstruction.
Holes of 1 Mpc and smaller are less important; these are not biasing
the density field at $k=0.1\ihmpc$ much.  
Of course, the reconstruction techniques themselves must
be able to deal with this mildly gappy data.
Gaps are less of an
issue at $z\gtrsim2$ where reconstruction is not a large advantage.

Photometric redshift surveys for acoustic oscillations will not measure
the 3-dimensional density field on the scales required to do density-field
reconstruction.  This is not a problem at $z>2$, where the real-space
density field is not degraded by non-linearities much anyways, but is
a disadvantage relative to spectroscopic surveys at $z<2$.

Much of the work on reconstruction has been focused on constructing the
velocity field from the density field, or vice versa \citep{nusser_etal_1991}.  
We are actually
interested in constructing the displacement field from the density field.
In linear theory, the velocity and displacement fields are the same, but
differences enter at higher order in perturbation theory.  We expect that
much of the work on velocity-field reconstruction will be applicable
to the displacement-field problem, and we note that in detail the
displacement field may be somewhat easier to infer, as virialization affects the
velocity field substantially but the displacement field very little.

The galaxies used in large-volume baryon acoustic oscillation surveys
are typically biased, and it will be important to build reconstruction
techniques that can handle this bias.  Again, one must remember that 
only modest performance is required of the reconstruction; if one can
fix half of the displacements, one will have achieved most of the gains.
All that is required is that light trace mass reasonably well on large
scales.  This is well established by the simple behavior of bias 
\citep[e.g.][]{Teg04},
the detection of large-scale redshift distortions \citep[e.g.][]{Pea01},
and the galaxy-mass cross-correlation measured from weak lensing
\citep{She03}.
The errors that enter from imperfect knowledge of the underlying 
cosmology or exact level of bias (e.g., the value of the redshift
distortion factor $\beta$) are much smaller than the required reconstruction
precision.


Finally, while reconstruction can improve the precision of the acoustic
scale measurement, one must also verify that the measurement remains
unbiased.  This will require further work, as one must also decide in
detail how one is measuring the distance information from the observed
clustering, particularly in the case of an anisotropic clustering
analysis.  One would expect reconstruction to at least partially fix
any bias that enters from non-linear gravity, but this has not yet
been studied at the sub-percent level.

In conclusion, we believe that density-field reconstruction offers 
a significant opportunity to improve the baryon acoustic peak method
by a factor of 2 to 3, particularly at lower redshifts.


\bigskip
We thank Martin White for useful conversations.
DJE and HS are supported by grant AST-0407200 from the 
National Science Foundation.  DJE is further supported by
an Alfred P.\ Sloan Research Fellowship. DNS and ES are
partially supported by NASA ATP grant  NNG04GK55G.
ES is also supported as a PICASSO fellow through the NSF IGERT program.



\end{document}